# The Kapitza - Dirac effect.

H. Batelaan [1]

Abstract. The Kapitza - Dirac effect is the diffraction of a well - collimated particle beam by a standing wave of light. Why is this interesting? Comparing this situation to the introductory physics textbook example of diffraction of a laser beam by a grating, the particle beam plays the role of the incoming wave and the standing light wave the role of the material grating, highlighting particle - wave duality. Apart from representing such a beautiful example of particle - wave duality, the diffracted particle beams are coherent. This allows the construction of matter interferometers and explains why the Kapitza - Dirac effect is one of the workhorses in the field of atom optics. Atom optics concerns the manipulation of atomic waves in ways analogous to the manipulation of light waves with optical elements. The excitement and activity in this new field of physics stems for a part from the realisation that the shorter de Broglie wavelengths of matter waves allow ultimate sensitivities for diffractive and interferometric experiments that in principle would far exceed their optical analogues. Not only is the Kapitza - Dirac effect an important enabling tool for this field of physics, but diffraction peaks have never been observed for electrons, for which is was originally proposed in 1933. Why has this not been observed? What is the relation between the interaction of laser light with electrons and the interaction of laser light with atoms, or in other words what is the relation between the ponderomotive potential and the lightshift potential? Would it be possible to build interferometers using the Kapitza - Dirac effect for other particles? These questions will be addressed in this paper.

## 1. Introduction

In 1933 Kapitza and Dirac suggested that electrons could diffract from a standing light wave [1] (figure 1). A demonstration of this effect would elegantly illustrate particle - wave duality. From figure 1 it is clear that the effect proposed is the equivalent of von Laue scattering as observed first for neutrons penetrating a crystal, where the grating is formed by the periodic structure of the atoms in a crystal. [Insert figure 1 about here] Both von Laue scattering and (reflective) Bragg scattering can be observed for thick crystals in contrast to observation of diffractive scattering for thin crystals [2]. We will refer to experimental conditions leading to the first two examples with the Bragg regime and to experimental conditions leading to the latter example with the diffractive regime.

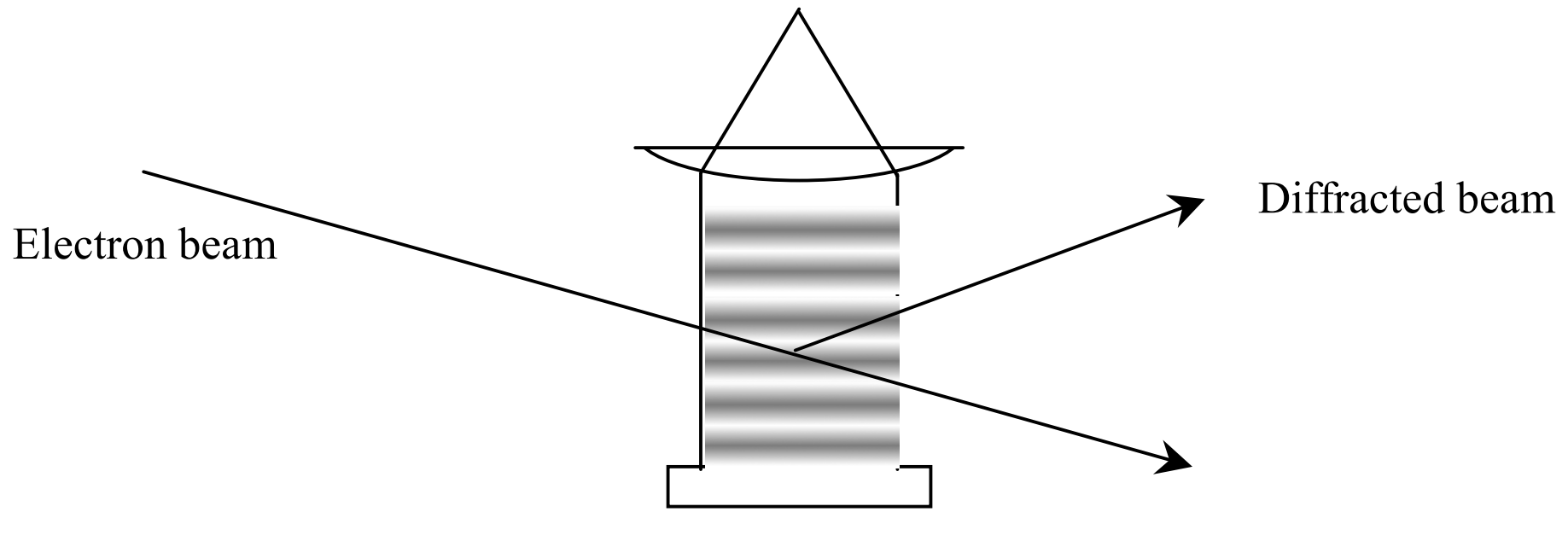

**Figure 1. Adapted from Kapitza and Dirac's original paper. Electrons diffract from a standing wave of light formed by retro-reflecting collimated light from a mirror. The electrons form a wave while the light plays the role of a grating.**

[1] Authors' address: Behlen Laboratory, University of Nebraska-Lincoln, Lincoln NE 68588-0111, USA. email: hbatelaa@unlserve.unl.edu

Kapitza and Dirac estimated that the relative strength of the deflected beam relative to the straight through going beam would be $10^{-14}$ using a mercury arc lamp. To obtain a useful 50/50 beam splitter much higher light intensities are needed. For this reason it is clear that attempts to measure the Kapitza - Dirac effect had to wait for the development of the laser. Shortly after the development of the laser attempts were made [3-6] to observe the KD effect. Some controversy seems to exist in the literature concerning its observation. Although early investigations reported an effect (some followed by retractions), recent reviews state that the KD - effect has not been observed for electrons [7, 8]. In the sense that separate diffraction peaks have not been observed this unfortunately appears to be correct. How unfortunate this is becomes especially clear when we realise that the beams emerging from a grating are coherent with each other. This means that Kapitza and Dirac might have called their paper: 'A coherent beam - splitter for matter waves'. In optics a simple piece of coated glass constitutes nothing less then a coherent beam - splitter. Such a simple optical element is however extremely important since it is the centrepiece of interferometers. Because no partially reflecting 'piece of glass' exists for atoms, ions, neutrons or electrons, a natural choice to produce coherent beams is a grating. Material gratings consisting of a crystalline material have been used long before standing waves were used as 'light crystals'. For neutrons, which can penetrate matter, and which have a short de Broglie wavelength at typical available energies, material crystals are a natural choice. For atoms both micro - machined material gratings and gratings formed from a standing wave of light have been realised [9] and have differing advantages. For charged particles, particularly at low energies, a standing wave of light may arguably be the natural choice. Even a small magnetic or electric effect caused by the presence of material would cause large deflections and smear out any sharp diffraction peaks. A major application of the KD effect thus concerns the field of matter optics.

The excitement this possibility holds is the short de Broglie wavelengths of matter waves and the promise that these short wavelengths yield a sensitivity for interferometric experiments that far exceed their optical analogues. Gravitational acceleration measurements can now be performed with an atom interferometer with an impressive relative accuracy of about $10^{-9}$ [10]. The best atom interferometer for detecting rotation [11] is now for the first time more sensitive than the best optical device.

## 2. Atoms

Following a suggestion by Altshuler *et al.* [12], the first observation of the Kapitza - Dirac effect for atoms [13, 14] was facilitated by the resonant enhancement of the interaction between *atomic* electrons and light (figure 2).

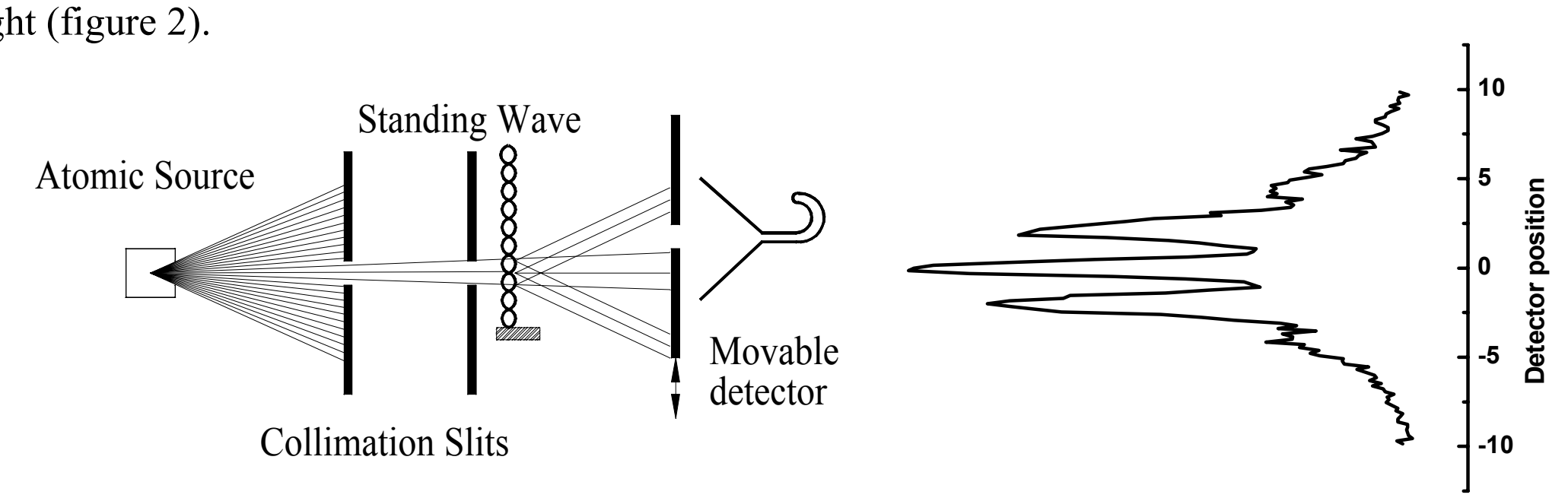

**Figure 2. (Left) A schematic setup for the Kapitza-Dirac experiment. (Right) Data for atom diffraction from a grating of 'light' taken at the University of Innsbruck [15]. The diffraction peaks in the number of detected particles are separated by two photon recoil momenta each.**

Passing a well - collimated atomic beam through a standing wave laser beam can be described accurately by an atomic plane wave accumulating a sinusoidal phase shift as a function of position. A simple thermal source, where the atoms are not coherent with each other can be used. This is not in conflict with the description of the atoms in terms of a plane wave, because this is a one - particle experiment, where effectively only one atom at a time propagates through the apparatus. [Insert figure 2 about here] The phase shift accumulated by the atomic wave is caused by the interaction $-\frac{1}{2}\alpha E^2$ between the induced atomic dipole, $d = \alpha E$, and the electric field $E$ associated with the standing wave, where $\alpha$ is the polarisability. The quadratic dependence on the electric field ensures that the interaction does not average out to zero (but to one half) for the fast oscillating electric field associated with a laser. The phase shift accumulated by the atomic wave follows the spatial periodicity of the laser field and is called a phase grating, or sinusoidal lightshift potential. This may be contrasted with absorption gratings where part of the atomic beam is removed.

Alternatively, we can adopt the view that the atom is interacting with photons. First, an atom absorbs a photon from the laser beam, and subsequently the atom is stimulated to emit a photon by the retro - reflected laser beam, thereby suffering a net momentum change of two photon recoils $2\hbar k$. Because the atom enters and leaves the interaction region in the ground state the atomic energy (kinetic and internal) does not change. This view affords a simple pictorial presentation of the Bragg and diffractive regimes (figure 3). [Insert figure 3 about here] In the first experimental report on Bragg scattering [13] the boundary between Bragg scattering and diffractive scattering was given in terms of the uncertainty $\Delta\phi$ in the direction of the absorbed photon (figure 3 left). This gives an uncertainty of momentum $\hbar k\,\Delta\phi$. The uncertainty in position where the photon will be absorbed, i.e. somewhere in the laser beam with waist, $\Delta w$, diffraction limits the value of $\Delta\phi$. This allows for the conservation of momentum and kinetic energy in the scattering process. When the value of $\Delta\phi$ is much larger than the angle, $\lambda_{dB}/\lambda_{opt}$, between the diffracted orders the atom can be scattered into many orders. This is the diffractive (or is sometimes called Raman - Nath) regime. In the Bragg regime the uncertainty in the photon angle $\Delta\phi$ is much smaller than the angle between the diffracted orders, thus the atom can not interact with two photons unless a special incident angle is chosen such that the momentum is conserved. This condition is satisfied at the Bragg angle.

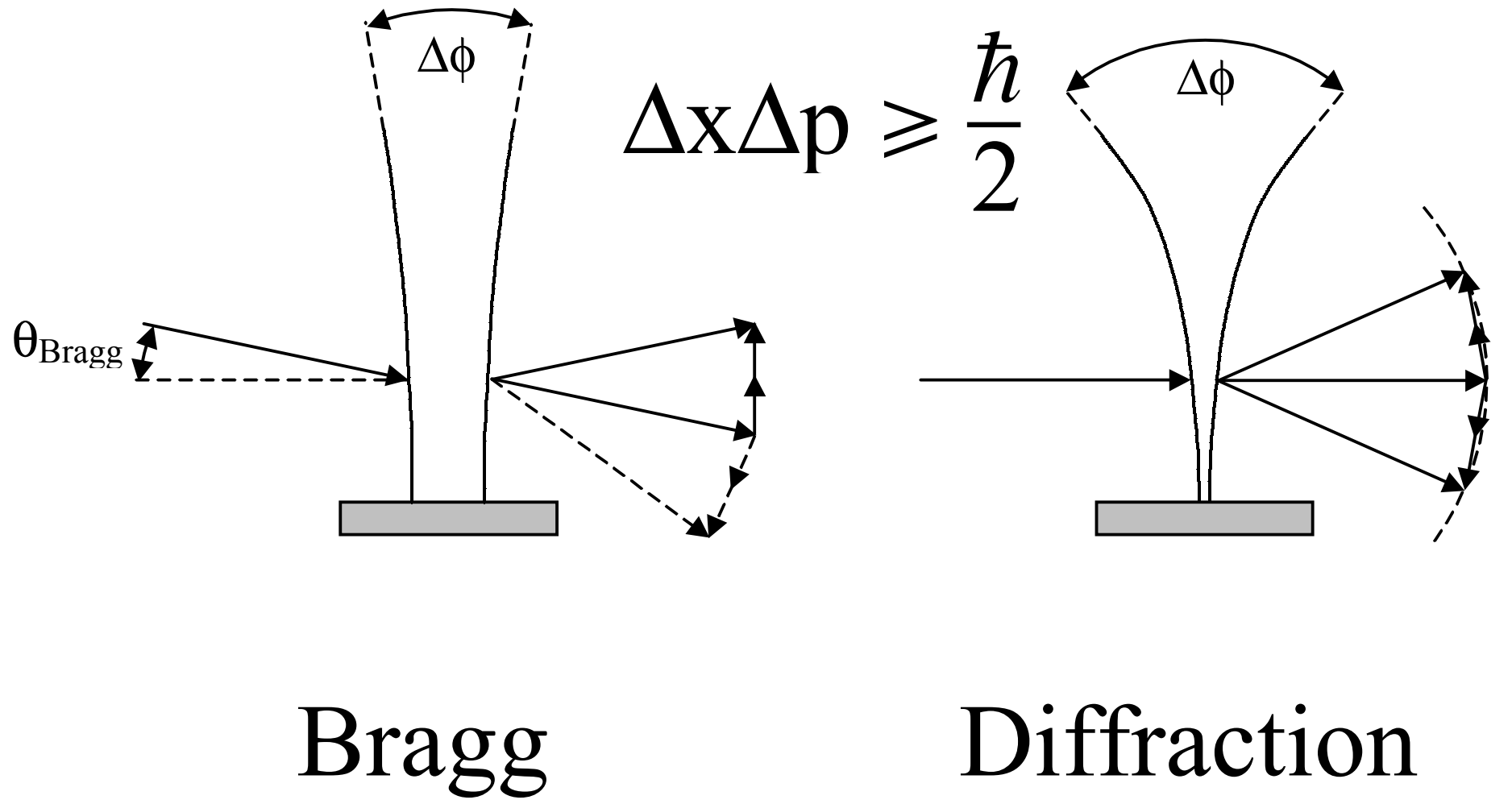

**Figure 3. Atoms moving through a standing light wave can be Bragg scattered (left) or diffracted (right). Absorption and subsequent stimulated emission only allow change of momentum of two photon recoils. Both effects can be described pictorially based on the position-momentum uncertainty principle. For a wide light wave the large uncertainty in the interaction position, $\Delta x$, leads to a small uncertainty in the momentum, $\hbar k\Delta\phi$, and only at the Bragg angle, $\theta_{Bragg}$, can energy and momentum be conserved (left). For a tightly focussed light wave many diffraction orders can be reached (right), while the incident angle is not critical.**

Another view can be adopted which is based on the energy - time uncertainty relation (figure 4). [Insert figure 4 about here]. This view is based on a 1 - D argument and must be used with care[2]. This view is also very helpful in that it is closely related to 1 - D simulations and can easily be linked to the experimental conditions. During the interaction time, $\Delta t = \Delta w / v$, an atom samples the frequency of a photon with an uncertainty of $1/\Delta t$. This corresponds to an uncertainty in the amount of energy absorbed by the photons of $\Delta E = \hbar/(2\Delta t)$. Energy conservation is satisfied by emitting and absorbing photons of different frequency from the light field. When $\Delta E$ is much larger than the recoil shift of the atom,

[2] Note that the energy-time picture predicts that the atoms change their kinetic energy.

$\varepsilon = \hbar^2 k^2 / 2m$, associated with the absorption of a photon, many different diffraction orders can be reached (figure 4 right). This is called the diffractive regime. When ΔE is much smaller than the recoil shift only one other diffraction order can be reached, and only then when the atom enters the interaction region at the Bragg angle. This is called the Bragg regime (figure 4 left). We note that the use of the energy - time uncertainty relation is often discouraged in view of the absence of a time operator from which the uncertainty relation could be derived. This problem is discussed extensively in the literature. Suffice it to say that for a two level - system an effective energy - time uncertainty relation can be defined [16, 17], and our system can be reduced to two level - systems coupled by two photon transitions.

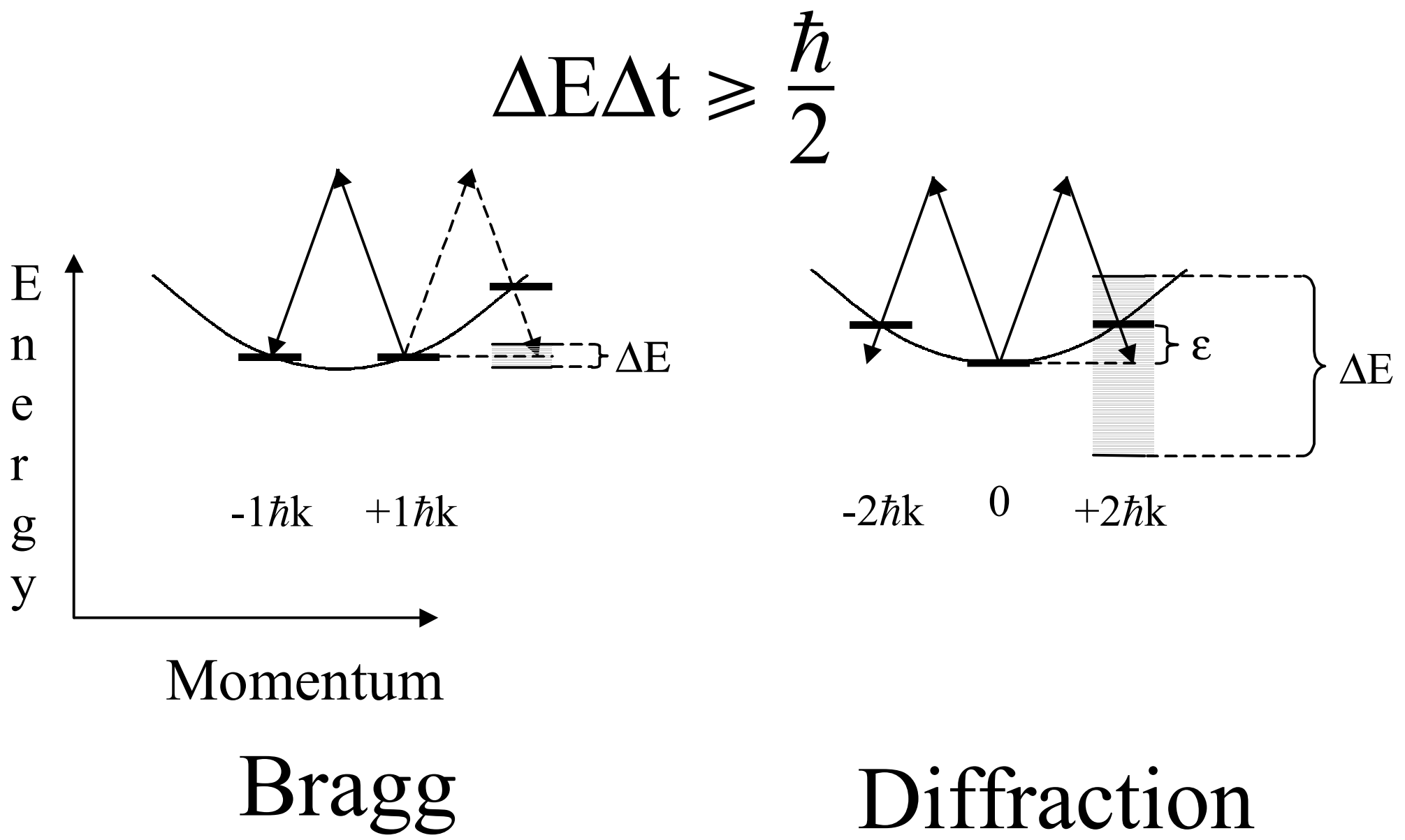

**Figure 4. Atoms moving through a standing light wave can be Bragg scattered (left) or diffracted (right). Both effects can be described pictorially based on the energy-time uncertainty principle. For a wide light wave the large uncertainty in the exact moment of interaction, Δt, results a smaller uncertainty in transferred energy than the energy shift associated with a photon recoil (left). This is why energy and momentum can only be conserved when the atoms enter the light wave with 1 ħk of momentum, which corresponds to the Bragg angle. For a short interaction time many diffraction orders can be reached.**

It is important to realise that both the above arguments, based on either the position - momentum or energy - time uncertainty relation, are a necessary but not a sufficient condition to distinguish between the two regimes. For the former case, in a laser beam with a Gaussian beam profile some photons with a large incident angle will be present. For the latter case, in a laser beam with a certain laser linewidth some photons with a large energy will be present. If enough photons are present an off - resonant transition can still be driven. Thus, even though the kinetic energy shift for neighbouring diffraction orders may be larger than the uncertainty ΔE by choosing long interaction times, scattering into many diffraction orders may occur when high laser intensities are used. High laser intensity corresponds to strong lightshift potentials and to discriminate between Bragg and diffractive scattering, one needs to determine whether the lightshift potential, U, is larger than the recoil shift, ε, or not. Note that a recoil shift is equal to the transverse kinetic energy of an atom incident at the Bragg angle. So the condition $U < \varepsilon$ means that an atom incident at the Bragg angle will pass over the potential hills. One can for such a weak potential (at sufficiently long interaction times) expect a transmitted part and a reflected part of the atomic beam. These are the two Bragg peaks. When U is much larger than the recoil shift we are in the diffractive regime.

The two conditions mentioned above, the first for the interaction time and the second for the potential strength, correspond to two experimental parameters, laser beam width and laser intensity, which can be adjusted independently. Since we have two conditions, four regimes instead of two exist (figure 5). When U << ε and simultaneously h / Δt >> ε, the atom experiences a too small interaction for a too short time to be scattered at all. On the other hand, when h / Δt << ε and U >> ε the atom can be channelled in the wells of the lightshift potential. [Insert figure 5 about here] [Insert table 1 about here] Based on the above discussion one finds that appreciable Bragg scattering or diffraction will occur when the value of the critical parameter U Δt is of the order of unity.

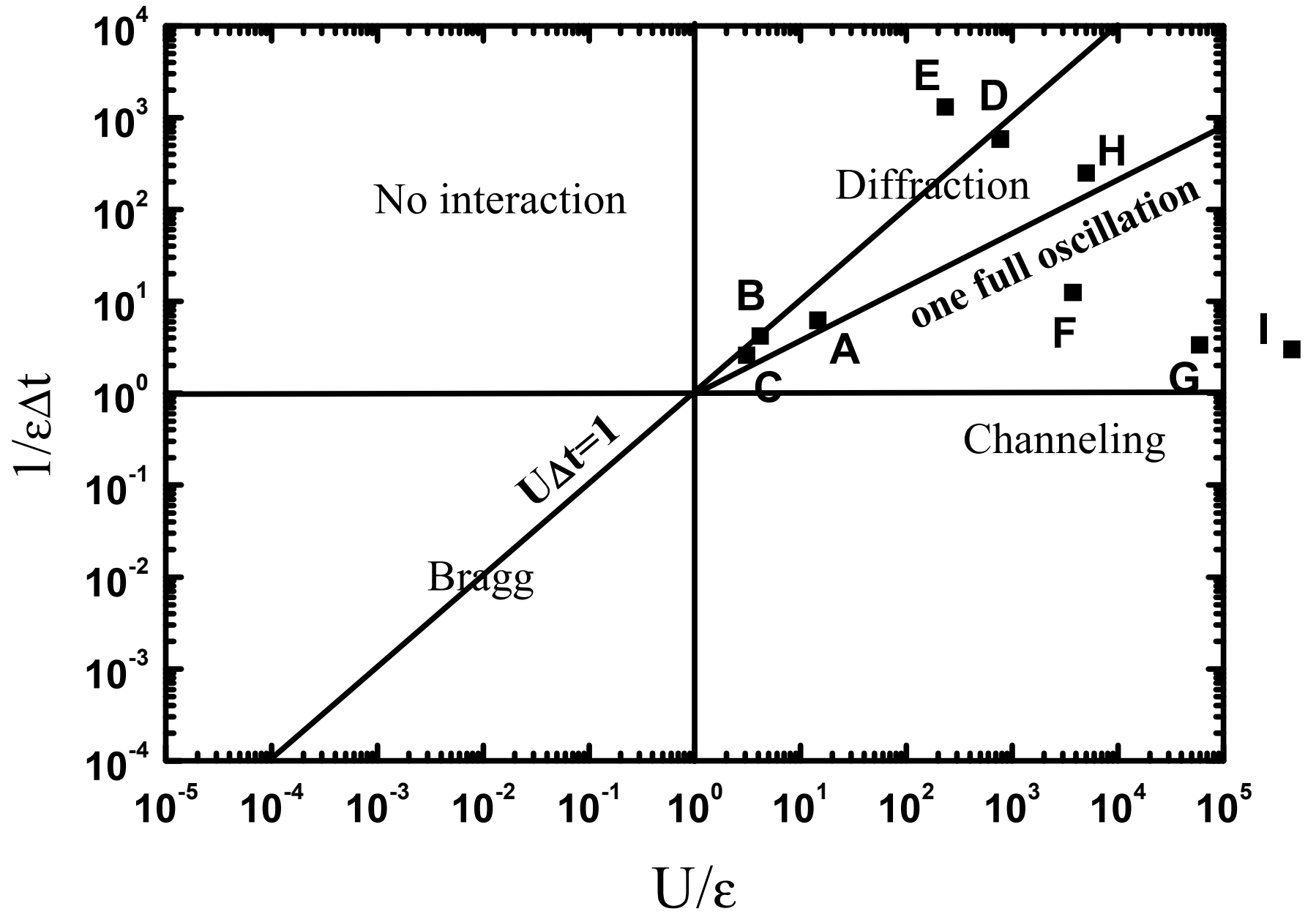

**Figure 5. The parameter space for atom photon scattering with some selected experimental results for atoms at thermal velocities. The references for the indicated points are given in table 1. In short, points E and D show symmetric scattering with respect to the incoming beam indicative of diffraction. Points A,B and C show Bragg scattering, F and G channeling, I guiding with a hollow fiber, and H lithography.**

| Reference | Atom | U (MHz) | 1/Δt (MHz) | ε (KHz) |
|---|---|---|---|---|
| A [14] | Na | 0.35 | 0.15 | 24 |
| B [18] | Ne | 0.10 | 0.1 | 24 |
| C [19] | Ar | 0.023 | 0.02 | 7.5 |
| D [13] | Na | 18.6 | 14 | 24 |
| E [15] | Ar | 1.65 | 10 | 7.5 |
| F [20] | Cs | 45 | 0.150 | 12 |
| G [21] | Li | 2188 | 0.12 | 37 |
| H [22] | Cr | 100 | 5 | 20 |
| I [23] | Rb | 1500 | 0.01 | 3.5 |

**Table 1. Reported or estimated values of the parameters for the experiments indicated in figure 5 are given.**

It is interesting to plot the parameters U/ε and 1/(εΔt) for several experiments. Experiments A, B, C are examples where one diffracted beam and one straight through going beam are observed in accordance with Bragg scattering. In experiments D and E symmetric diffraction patterns around the direction of the incoming beam are observed as one would expect to find for diffraction (see also theory below). In experiments F and G the light intensity is sufficient to confine the atoms to the valleys of the standing light wave, while the interaction time is long compared to the classical oscillation frequency of the atoms in the potential valley. This regime traps atom in the potential valleys and is typically named channelling.

Limiting the interaction time to a quarter oscillation of the atom in the potential, as has been done in experiment H, creates a row of lenses separated by half the optical wavelength of the laser light. This technique has been used to form lines of chromium atoms with widths of tens of nanometers on a substrate. Research is being performed to reduce the width of the lines, which is limited by chromatic and spherical aberration [24]. Such a lithographic technique would be very exciting if it can be applied to form quantum lines and quantum dots of ferromagnetic materials in view of the discoveries of GMR and PMA [25]. Experiment I involves the guiding of atoms with a hollow fiber. The evanescent wave of the laser light guided through the walls of the fiber prevents the atoms from hitting the walls. With such techniques still useful higher values for the potential and the interaction time can be reached with no appreciable spontaneous emission.

## 3. The lightshift and ponderomotive potentials

As mentioned above, passing an atom with polarisability $\alpha$ through a standing wave laser beam can be described by the lightshift potential, $V_L = -\frac{1}{2}\alpha E^2$. Passing an electron through a standing wave laser beam can be described by the ponderomotive potential. At first glance it is surprising that for either case the rapid oscillation of the electric and magnetic fields in the laser beam do not average out to zero. The standard explanation for the lightshift potential is typically that the potential depends quadratically on the electric field strength, where the induced dipole and the dipole interaction depend linearly on the field strength. This energy picture may be somewhat unsatisfying since the dynamics of the atom in the rapidly oscillating lightfield is not addressed and the relation to the ponderomotive potential for electrons is unclear. A force picture provides a dynamical view of the atomic motion and allows the ponderomotive potential to be understood in equivalent terms. We start by considering the ponderomotive potential. First we need to discuss both the electric and magnetic fields in a standing light wave.

Using a standing wave solution of the Helmholtz equation given by the vector potential, $A_z = A_0 \cos kx \sin \omega t$, the expressions for the electric and magnetic field are

$$E_z = \frac{\partial}{\partial t} A_z = A_0 \omega \cos kx \cos \omega t; B_y = -\frac{\partial}{\partial z} A_x = A_0 k \sin kx \sin \omega t . \quad (1)$$

The electric field oscillates π/2 out of phase with the magnetic field both in space and time. With the help of figure 6 the spatial part can be visualised. [Insert figure 6 about here] An in - phase oscillation would result an average Poynting vector that is parallel or anti - parallel to the k - vector of the light. For equal intensity counter propagating light waves this is not appropriate.

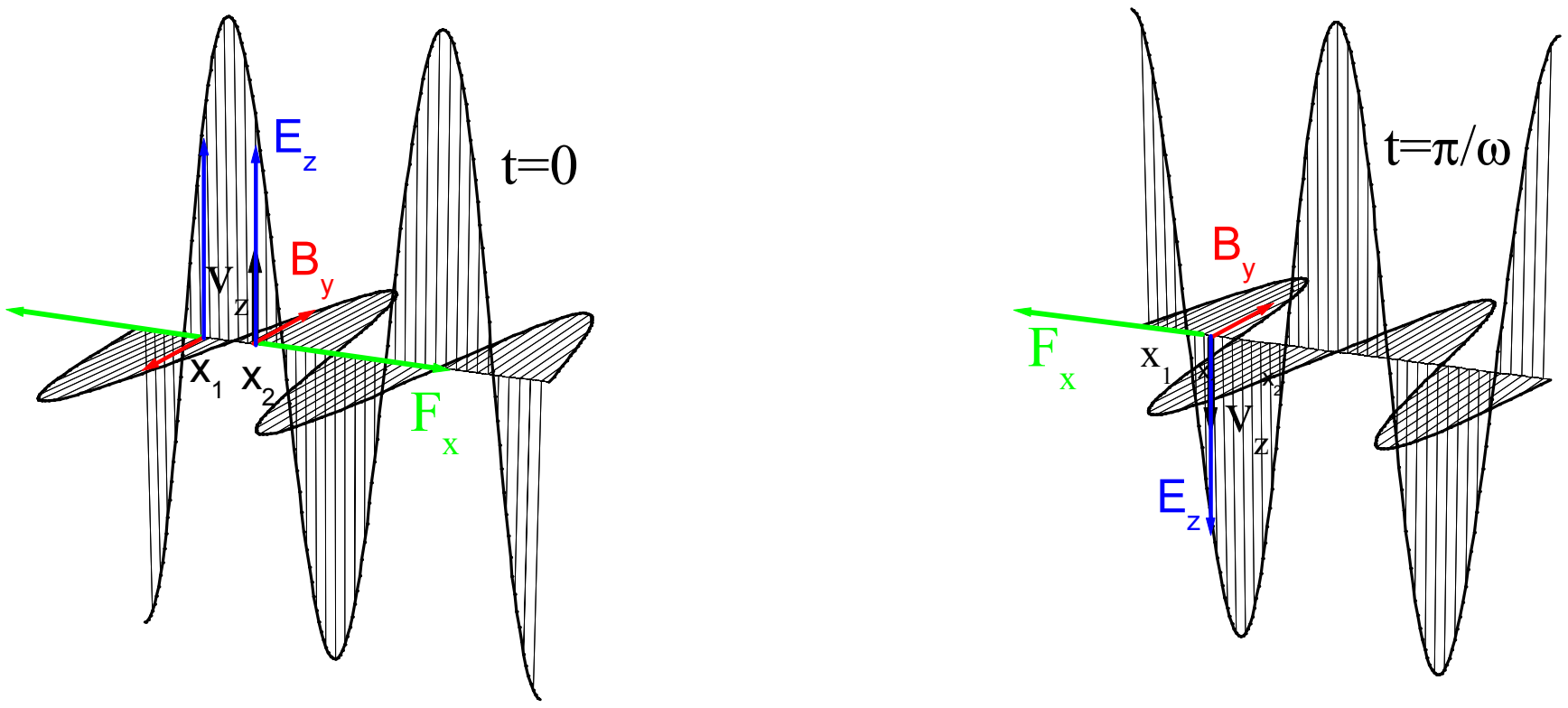

**Figure 6. The electric field $E_z$, the velocity $v_z$, the magnetic field $B_y$, and the Lorentz force $F_x$ are shown at two positions for a standing light wave (left) and at a time half a period later (right). After half a period of the laser field oscillation, both the electric and magnetic field flip sign causing the Lorentz force on a free electron not to average out to zero over time. The Lorentz force varies twice as fast with position as does the electric field.**

Consider an electron placed halfway between a maximum and a zero crossing in the electric field ($x=x_1$) (figure 6 left). The electron will be accelerated in the electric field direction (z - axis) and due to its acquired velocity experience a Lorentz force along the propagation direction of the light, $F_x = v_z B_y$. The electron velocity in the electric field direction will oscillate and lag behind by a phase of $\pi/2$ with respect to the electric field, due to its inertia,

$$\upsilon_z = \tfrac{eA_0}{m} \cos kx \sin \omega t \,. \tag{2}$$

This phase shift causes the velocity of the electron to oscillate in phase with the magnetic field and thus the Lorentz force will not average to zero over an oscillation period because both the electron's velocity and the magnetic field change sign simultaneously (figure 6 right). The electric force on the electron will average out to zero. The resultant force will thus be directed along the x - axis. For a more complete classical treatment of an electron in a standing light wave see for instance [26]. What is the position dependence of the Lorentz force along the x - axis? Consider an electron placed a quarter wave displaced from $x_1$ at $x=x_2$. Here the electric field has the same sign as at position $x=x_1$, but the sign of the magnetic field is reversed. Thus the direction of the force at $x_2$ is reversed with respect to the force at $x_1$. From the force we can obtain an effective potential; the ponderomotive potential

$$V_P = \tfrac{e^2 A_0{}^2}{4m} \cos^2 kx \,. \tag{3}$$

To get an experimentally more useful expression the relation between vector potential and laser intensity

$$I = \tfrac{1}{2} \varepsilon_0 c \omega^2 A_0{}^2 \tag{4}$$

can be used.

We now follow the same approach but for a charged mass attached to a spring. This represents a classical model for a polarisable atom. The final result of this discussion will be that the average Lorentz force for this system yields the lightshift potential (see also [27, 28]). The following features of the lightshift potential have to be explained, a) for an atom driven at its resonant frequency the lightshift potential is zero, b) going from above to below resonance the lightshift potential changes sign, and c) far from resonance the potential is small. On resonance the velocity of the mass is oscillating in phase with the driving external force. Consequently, the velocity of the charge is oscillating in phase with the electric field and is thus out of phase with the oscillating magnetic field. In contrast to the free electron case, this causes the Lorentz force to average out to zero. A little bit above resonance the velocity of the electron is slightly ahead in phase with respect to the electric field oscillation and the Lorentz force does not average out to zero. A little bit below resonance, the phase shift is opposite and again the Lorentz force does not average out to zero, but has an opposite sign. The analytic solution of Newton's equation for the velocity $\upsilon_z$ of a charge with mass on a spring is given by,

$$\upsilon_z = (-\omega_0 \tfrac{\sin \omega_0 t}{\omega^2 - \omega_0{}^2} + \omega \tfrac{eE}{m} \tfrac{\sin \omega t}{\omega^2 - \omega_0{}^2}) \cos kx \,, \tag{5}$$

where $\omega_0$ is the resonant frequency. The Lorentz force averaged over time, can be obtained by integration, and the effective lightshift potential follows,

$$V_L = \tfrac{1}{4} \frac{q^2 / m}{\omega^2 - \omega_0{}^2} E_0{}^2 \cos^2 kx \tag{6}$$

Defining the classical expression for the atomic polarisability appropriately as $\alpha = \frac{q^2/m}{\omega_0^2 - \omega^2}$ [29], this expression reduces (apart from a factor of two) to the same expression obtained from the simple energy argument, $V_L = -\frac{1}{4}\alpha E^2$ [3]. It is satisfying to see that the limit for a weak spring of the lightshift potential, when the electron is effectively free and $\omega_0 \to 0$, gives the ponderomotive potential. This is why it is sometimes stated the free electron has a resonance at $\omega_0 = 0$.

The above arguments may give the false impression that the consequences of the light shift and ponderomotive potential in real experiments are easy to predict. The above arguments are thought to be an interesting starting point for discussion and useful for rapid estimation of experimental parameters. The difficulties involved with for instance understanding multi - photon ionisation, where both potentials play a role, are notorious [30]. The difficulties with understanding the quantum aspects of the ponderomotive potential by itself are likewise large. An example of this may be that the Schrödinger equation is not satisfied by the non - relativistic limit of the solutions to the Dirac equation [31].

## 4. Diffraction theory

We would expect that the motion of a particle in a sinusoidal potential, such as the ones discussed in the previous section, should be treated quantum mechanically if the particle's de Broglie wavelength associated with the motion in the direction of the standing wave is comparable to the potential's period,

$$\lambda_{dB} = \lambda_{optical}/2. \tag{7}$$

For particles entering the standing wave at right angles and experiencing a momentum change of two photon recoils, $\Delta p = 2\hbar k$, this condition holds exactly. This means that when the interaction is just strong enough to give an appreciable effect, we may expect that a quantum mechanical treatment of motion is necessary. The description of our problem can be reduced to one dimension by ignoring the longitudinal motion. The photon recoil will only change the transverse velocity of the particles. For the Hamiltonian

$$H = -\frac{\hbar^2}{2m}\frac{\partial^2}{\partial x^2} + V_0 \cos^2 kx, \tag{7}$$

the Schrödinger equation can be solved using a trial solution of the form

$$\psi = \sum_n c_n e^{inkx}, \tag{8}$$

which describes the motion of the particle in terms of plane waves separated by one photon recoil each. With a minimum of manipulation, differential equations for the coefficients $c_n$ can be found, which turn out to express the amplitude for finding the particle in the $n/2$ - diffraction order,

$$i\frac{dc_n}{dt} = (\varepsilon n^2 + \frac{V_0}{2\hbar})c_n + \frac{V_0}{4\hbar}(c_{n-2} + c_{n+2}), \tag{9}$$

where $\varepsilon n^2 = \hbar k^2 n^2 / 2m$ is the kinetic energy (in units of $\hbar$) of each plane wave with momentum $n\hbar k$. This equation is the basis for some of the physical pictures presented above. From the second term on the right hand side of equation 9 it can be seen that the momentum of the particle can only change by plus or minus 2 momentum recoils. For atoms we have ignored excited states and would expect that atoms can only change their motion by absorption followed by stimulated emission to return to the ground state and thus exchanging two photon recoils. Spontaneous emission can be suppressed by tuning the laser light far off resonance. For electrons, the absorption of one photon by a free electron is not allowed. But two photon recoils can be exchanged between the field and a free electron by stimulated Compton scattering. The solution of equation (7) can be found analytically and it is interesting to consider it for two cases. When the particle does not have enough kinetic energy to move over the potential crests ($\varepsilon << V_0/\hbar$) we are in the diffractive regime, and the solution (which can be readily checked by direct insertion) is

$$c_n = i^{n/2} e^{-\frac{i}{\hbar}V_0 t} J_{n/2}(V_0 t/\hbar) \Rightarrow |c_n|^2 = J_{n/2}{}^2(V_0 t/\hbar). \tag{10}$$

[3] To obtain the correct factors of 2 care should be taken with the definition of electric field associated with that of either the travelling or the standing wave and the polarization of the light.

The Bessel function solutions express that a symmetric diffraction pattern can be observed, where $|c_n|^2$ is the detection probability of finding the particle in the *n* - th diffraction order. Conversely, when the particle can easily move over the potential crests ($\varepsilon >> V_0/\hbar$) we are in the Bragg regime. Here, as discussed above, only two diffraction orders couple,

$$\begin{cases} c_1 = e^{-i\varepsilon t}\cos(V_0 t/4\hbar) \\ c_{-1} = -ie^{-i\varepsilon t}\sin(V_0 t/4\hbar) \end{cases} \Rightarrow \begin{cases} |c_1|^2 = \cos^2(V_0 t/4\hbar) \\ |c_{-1}|^2 = \sin^2(V_0 t/4\hbar) \end{cases} \tag{11}$$

where the probability for finding a particle in either diffracted particle beam 'pendulates' back and forth, as expressed by the so called 'Pendellösung' (11). This oscillation can be observed both as a function of laser intensity or interaction time. If one would like to obtain appreciable diffraction or Bragg scattering it can be seen from (10) and (11) that the parameter $V_0 t/\hbar$ should be of the order of one in agreement with our earlier qualitative discussion (figure 5). In other words, to find out if a certain laser beam can diffract a certain charged or polarisable particle then a good first step is to calculate the value of the parameter $V_0 t/\hbar$ using the lightshift potential $V_L$ (6) or the ponderomotive potential $V_P$ (3) (see experimental parameters section below for some suggestions). Numerically integrating eq (9) gives the probability of observing diffraction as a function of laser intensity and interaction time for somewhat more arbitrary experimental parameters and allows one to assess how rigorously for instance the condition $\varepsilon << V_0/\hbar$ should satisfied to enter the Bragg regime. Solutions close to the analytical ones (figure 7) can be readily obtained. A narrow interaction region results in electron diffraction (figure 7 left), a wide interaction region produces Bragg scattering (figure 7 right). [Insert figure 7 about here]

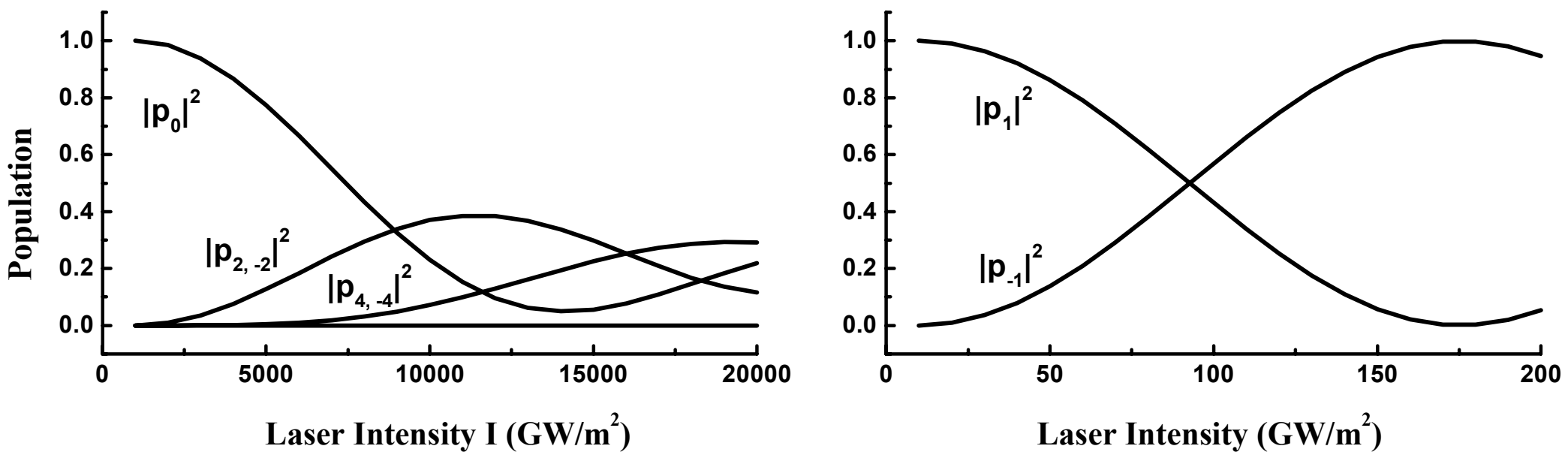

**Figure 7. The results of a quantum mechanical calculation showing diffractive scattering for an electron energy, *E*, of 10 eV and a laser beam waist, *w*, of 0.005 cm (left). Solutions closely resembling Besselfunctions are found (equation 10). For a much wider laser beam waist, *w*=0.5 cm, Bragg scattering shows a sinusoidal behaviour (equation 11). The relative strength of scattering for electrons remaining in the initial beam is indicated by $|p_0|^2$ (left) and $|p_1|^2$ (right). Scattering to the higher order diffraction peaks (the subscripts indicate the number of photon recoil momenta) is also indicated.**

What happens when the interaction time or the potential strength is increased? Assuming that the interaction is strong enough for the particle to scatter many photons then one can resort to a classical trajectory calculation, an example of which is given in figure 8. Particles moving through the trough of the standing light wave experience to a good approximation a parabolic potential. In this way each standing wave forms a small micro - lens. This idea has been used to deposit lines of atoms on a substrate and is thus a form of atomic lithography [22]. If the number of particles as a function of position is observed far away from the light wave then two peaks are observed corresponding to the maximum deflection angles. The maximum deflection angles occur at the maximum force and are thus attributed to the inflection points of the potential. This is a well - known phenomenon that occurs for a variety of physical systems and is called rainbow - scattering. This classical effect has been observed both for atoms [20] for which is has been called channelling and this effect has been observed for electrons [32] (figure 8 right) for which it has been called the high intensity Kapitza - Dirac effect. [Insert figure 8 about here]

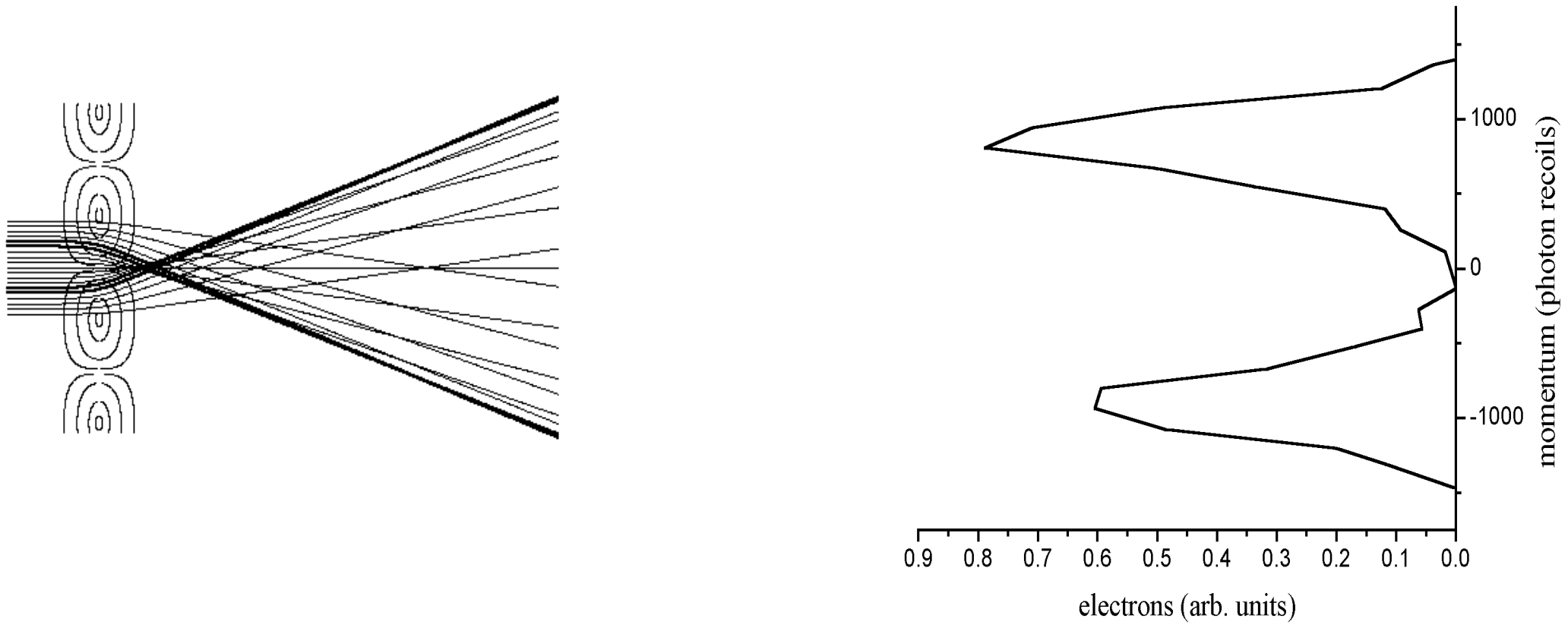

**Figure 8. Classical trajectories for particles moving through a sinusoidal potential (left). The particles are focussed just after they passed through the potential and the maximum deflection angles far away from the potential can be seen. Experimental data for electrons [33] show the rainbow peaks associated with the maximum deflection angles (right). In this experiment the electrons were produced in the laser light field by multi-photon ionisation and the observed splitting has thus also been referred to as a half process [34].**

The elementary discussion given above may (hopefully) lead the reader to ponder many questions. Would not spontaneous Compton scattering (a one-photon process) overwhelm stimulated Compton scattering (a two-photon process)? What happens when the laser intensity is increased and for example the classical oscillation frequency of the electron in the potential becomes equal to the laser light frequency? How does the magnetic field associated with the laser light influence the electron spin? The answers to such questions are beyond the scope of the present paper, but can be found in beautiful papers by Eberly [35], Bucksbaum [33], Guo *et al.* [34], and Chaloupka and Meyerhofer [36].

## 5. Suggested experimental parameters

Three decades ago, attempts to observe the Kapitza - Dirac effect for electrons failed to produce diffraction peaks [7, 8]. However, in view of the data shown in figure 8 (right) [33] strong enough standing waves can be produced to scatter about a 1000 photons. In the early experiments Bartell [3] and Schwartz [4] reported the observation of deflection of electrons with laser light, while others observed no effect [5, 6]. Why did these experiments fail to produce diffraction peaks? Fedorov [8] has suggested that an adiabatic turn - on is the main reason for the previous failure to observe the effect. This suggestion can be understood by considering the width of the incoming electron beam. In the Bragg regime electrons entering the laser beam at the Bragg angle can be diffracted. When electrons enter the laser beam at an angle slightly different from the Bragg angle, their diffraction depends on the time the electron spends in the laser beam. For longer interaction times the Bragg angle needs to be matched better to observe any effect. In the limit of infinite interaction times no electrons will be diffracted at all. So definitely a well - collimated electron beam is required. An experimental study could settle this issue.

What are the experimental conditions for which the Kapitza - Dirac effect might be observed? Some suggested experimental parameters are given in table 2 and 3, where $v$ is the velocity of the incoming electrons, $w$ is the laser beam width (the height is chosen to be 1 mm), $I$ is the intensity of the laser beam, and $\lambda$ is the wavelength of the laser light. These parameters are included to show that tabletop experiments are feasible.

| | Lifetime (s) | $\lambda_L$ (nm) | $\lambda_{atom}$ (nm) | $I$ (W/m$^2$) | $U^*\tau_{int}$ |
|---|---|---|---|---|---|
| Na | Ground state | 488 | (7lines) | $10^7$ | 0.4 |
| $Ar^*$ | 60 | 488 | (15 lines) | $10^7$ | 0.4 |
| $Ca^+$ | Ground state | 488 | 393 | $10^7$ | 0.1 |
| $Li^+$ | 49 | 488 | 548.5 | $10^7$ | 0.1 |
| $Ba^+$ | Ground state | 488 | 493 and 455 | $10^7$ | 2.5 |

**Table 2. Parameters for high intensity atom and ion diffraction. A value of the product of the lightshift potential U in units of Hz and the interaction time $\tau_{int}$ in the order of unity indicates that the number of diffracted particles would be a substantial fraction of the total number of incoming particles. The chosen intensity I and laser wavelength $\lambda_L$ corresponds to the use of a 1 W single line Argon ion laser beam focused to a width of 100 μm and a height of 1 mm. For all five species the lightshifts of the dominating lines have been added taking into account the relative sign of the lightshift.**

| | $\lambda$ (nm) | $I$ (GW/m$^2$) | $V_P\Delta t/\hbar$ (Hz) | $v$ (m/s) | $w$ (cm) | $V_P\Delta t/\hbar$ (Hz) |
|---|---|---|---|---|---|---|
| Bragg | 1064 | $10^2$ | $10^9$ | $2\times10^6$ | 0.5 | 2 |
| Diffractive | 1064 | $10^4$ | $10^{11}$ | $2\times10^6$ | 0.005 | 2 |

**Table 3. Proposed experimental parameters for electron diffraction.**

[Insert table 2about here][Insert table 3 about here] The resulting ponderomotive potential strength $V_P/\hbar$ and its product with the interaction time, $V_P\Delta t/\hbar$ is given. This value is a measure for the coupling strength between adjacent motional states and should of the order of unity to allow observation of appreciable diffraction peaks (Table 1 and above discussion). For atoms it is possible to perform experiments close to resonance with tens of mW of laser light. An example of our data is given in figure 9.

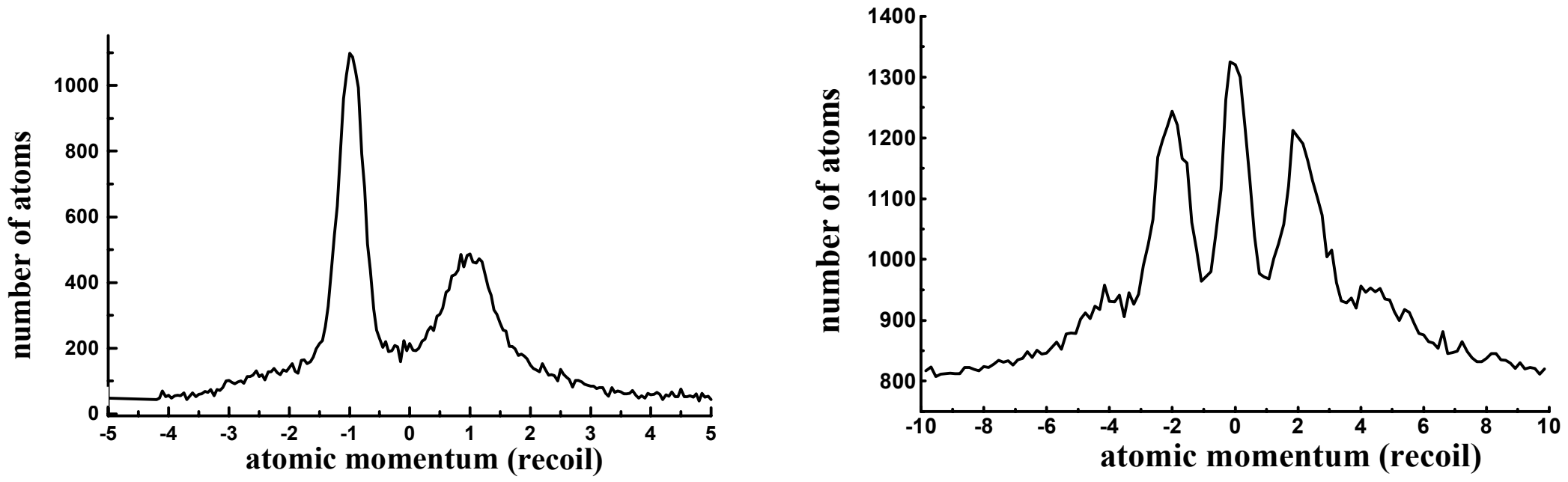

**Figure 9. Experimental data for atomic Bragg (left) and diffractive scattering (right) .**

[Insert figure 9 about here] Such an experiment requires a laser locked to resonance, which means that for each atomic species a different laser system is needed. With that in mind it is interesting to calculate the laser intensity needed to diffract atoms at say detunings of nanometers or more. Here an estimation of the interaction strength is complicated by the presence of the multitude of resonances that are present for any real atom. We can extend this discussion to ions. To our knowledge laser light has never been used to diffract an ion beam. For ions a complication is that both the ponderomotive potential (for charged particles) and the lightshift potential (for polarisable particles) play a role. Limiting ourselves to the lightshift potential, the results of our previous discussion and by using tabulated resonance lines [37] some particles that might be diffracted with laser light can be found (table 2). After our discussion showing that table-top experiments for different particles can be attempted we now turn our attention to the interesting physical problems that may be addressed with the realizationof such techniques.

## 6. Interferometry

"If a particle can reach a detector by at least two indistinguishable paths then the probability of detecting this particle will be given by the square of the sum of the amplitudes associated with each path'. Depending on the phase difference between the amplitudes of each path, differing amounts of interference, ranging from destructive to constructive interference, may occur. This is the basic working principle for any interferometer. Why is an interferometer such a widely used device? Two important characteristics of an interferometer are that a) any interaction causing a phase shift in one of the interferometer arms can in principle be detected, and b) the wavelength of the light or particles used for interferometers is short. The first characteristic ensures that one can use an interferometer to study a wide variety of effects, while the second ensures that one can do that with high accuracy.

### *6.1. Atom interferometry*

Both for electrons and atoms, Bragg scattering and diffraction split the incoming beam into two or more coherent parts (figure 9). Such a beam splitter can be used to build a Mach - Zehnder interferometer for a matter wave with a wavelength of less than $10^{-9}$ m, a thousand times less than the wavelength for a light interferometer (figure 10). [Insert figure 10 about here]

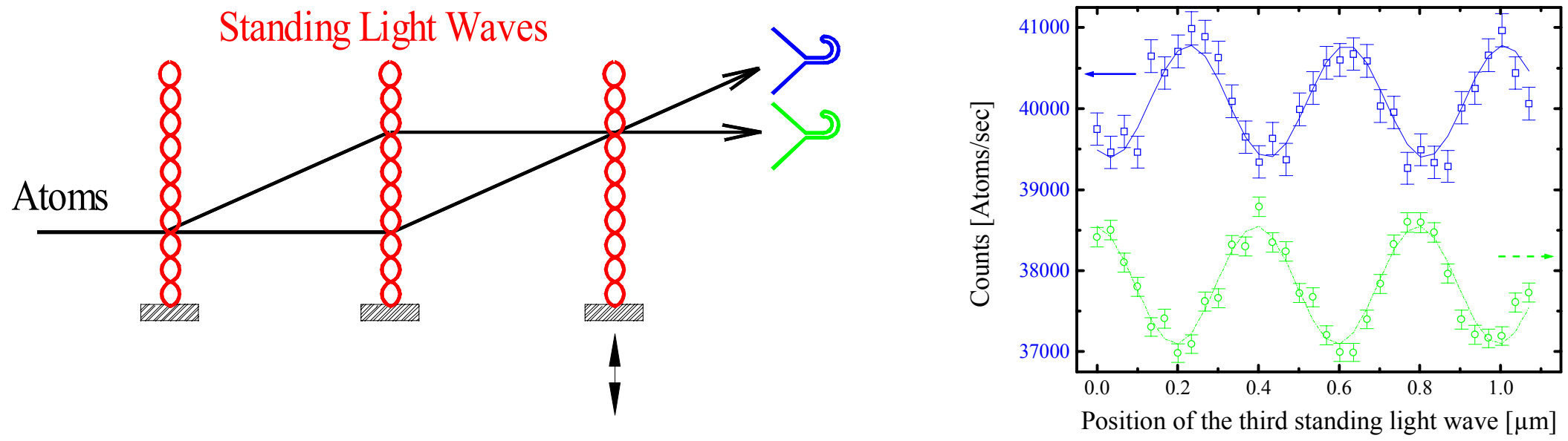

**Figure 10. Schematic of our Mach-Zehnder atom interferometer with gratings of light and two fringe patterns measured at the output beams 1 and 2 showing particle number conservation** [15]. **Three consecutive gratings were used the split and recombine an atomic beam. The light frequency was chosen close to resonance.**

What experiments can be performed with a Mach - Zehnder interferometer? In the following part of this paper we do <u>not</u> attempt to give a representative review of the vast amount of work done in the field of matter inteferometry. The goal is to give a feel for the sensitivity interferometers may have and to list some beautiful ideas available in the literature that might be addressed with such interferometers. A selection has been made to mention those ideas that may benefit from the use of the light gratings (i.e. the KD - effect).

### *6.2. Sensitivity*

Consider a Mach - Zehnder interferometer that rotates around an axis normal to the surface enclosed by the two interferometer arms. In one arm of the interferometer the wave tries to catch up with the rotation, while in the other the wave travels into the direction of rotation. The faster the rotation is, the larger the difference in number of waves becomes, when the two waves recombine. This rotation dependent phase shift of an interferometer's fringes is called the Sagnac effect and allows an interferometer to be used as a rotation sensor. What is the rotational sensitivity that may be expected for existing interferometers? In other words how accurately can one measure the rotation rate and how long would the measurement take? Following the approach of Oberthaler [38] the resolution $R$ is given by $R=k_g L^2/v$, where $k_g$ is the length of the reciprocal grating vector, $L$ is the distance between the gratings and $v$ is the particle velocity. The resolution is a measure for the amount of phase shift per unit of angular velocity. The sensitivity $S$ is given by $S=(R\,C\,n^{1/2})^{-1}$ , where $C$ is the contrast of the interference pattern and $n$ the number of particles detected per second. A sensitivity of $S=1\Omega_e\,s^{1/2}$, where $\Omega_e$ is the earth's rotation, means that the effect of the earth's rotation ($\Omega_e = 7\times10^{-5}$Hz/2π) could be measured in 1 second. The lower the value for S the more sensitive the interferometer is.

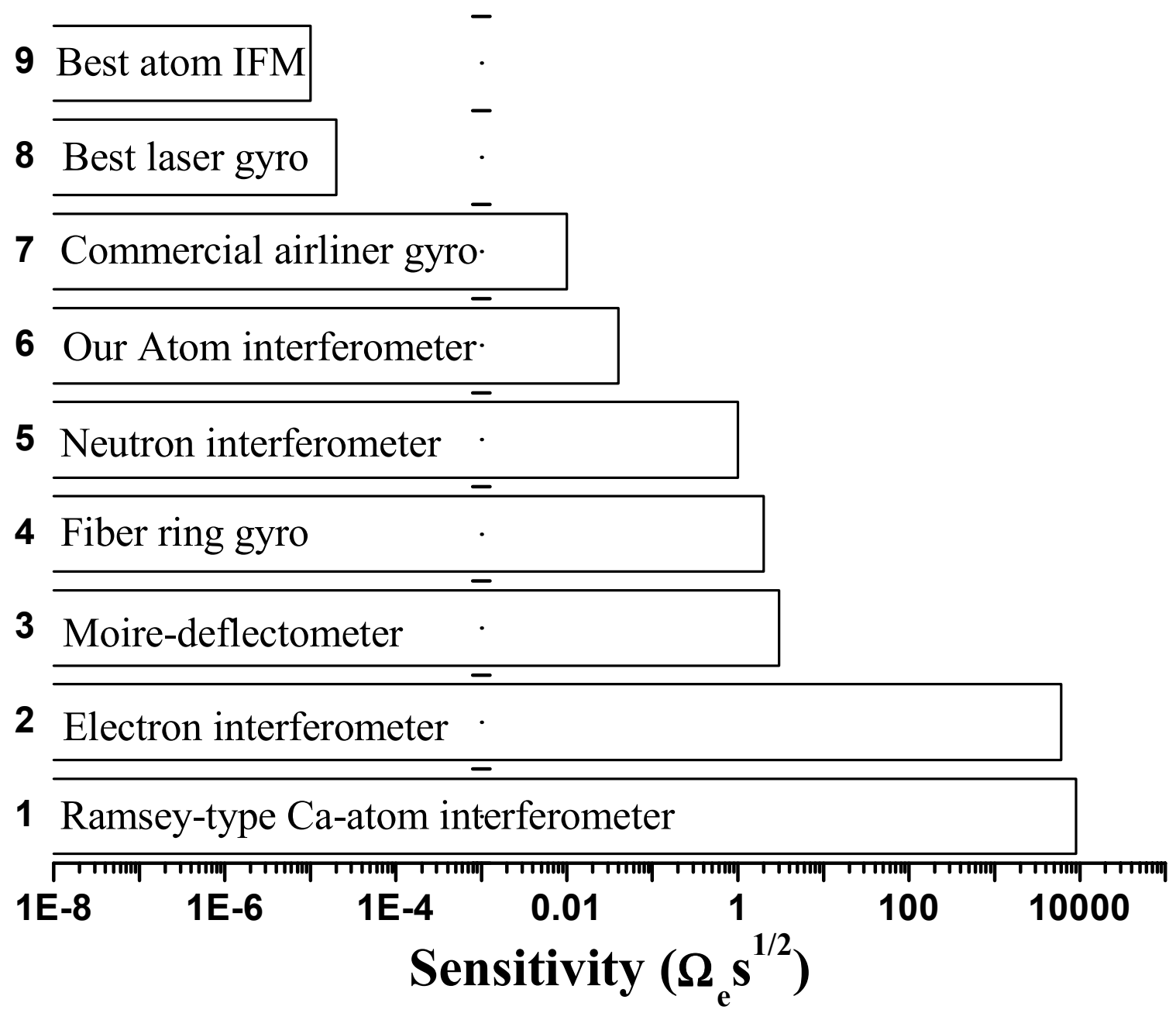

**Figure 11. Overview of rotation sensor sensitivity. A small value corresponds to a high sensitivity and is given in the measurable rotation rate (in units of the earth's rotation $\Omega_e$) in one second. The references for the devices 1-9 are [15, 38-45] , respectively.**

One would like to increase the resolution and lower the value for the sensitivity as much as possible. For a light interferometer, $v = c$, which would tend to reduce the sensitivity and lead to poor sensitivity. On the other hand the detection rate and the enclosed area of light interferometers can be very large, which explains the excellent sensitivity of light interferometers. For atoms the velocity is small and this helps the resolution, however the detection rate and enclosed area are typically relatively small. Experimentally, atom interferometers can now for the first time compete with the best light interferometers (figure 11). [Insert figure 11 about here] For electrons the velocity typically lies between that of light and atoms. For an electron interferometer with gratings of light the enclosed area would also lie between that of atoms and light. It is a matter of experimental investigation to determine what sensitivity could be reached using different particle types with Mach - Zehnder interferometers.

*6.3. Future directions*

*6.3.1 Amplification*

A natural idea to improve the sensitivity of a charged particle interferometer almost suggests itself. Can electrostatic lenses be used to obtain a large enclosed area by the interferometer arms? To deflect a finite sized beam a potential gradient is needed. This means that particles on one side of the beam pass through a different potential than particles passing on the other side of the beam, and thus accumulate a different phase, making it impossible to observe interference fringes. This basic problem might be reversed with appropriate control of additional electrostatic lenses [46]. In relation to the spin coherence in a Stern - Gerlach interferometer Bohm has stated that 'fantastic' control of the fields involved would be required [47]. It is an interesting question to establish how well one can experimentally control these phase shifts for the 'wave' coherence in a Mach - Zehnder interferometer. Relying on the close relation between optics and matter optics a simple experiment can be performed. The enclosed area of an optical Mach - Zehnder interferometer (with laser light passing through material gratings) may be enlarged by using cylindrical lenses. After the light has diffracted from the first grating a negative lens is used to increase the angle between the diffracted light beams (figure 12).

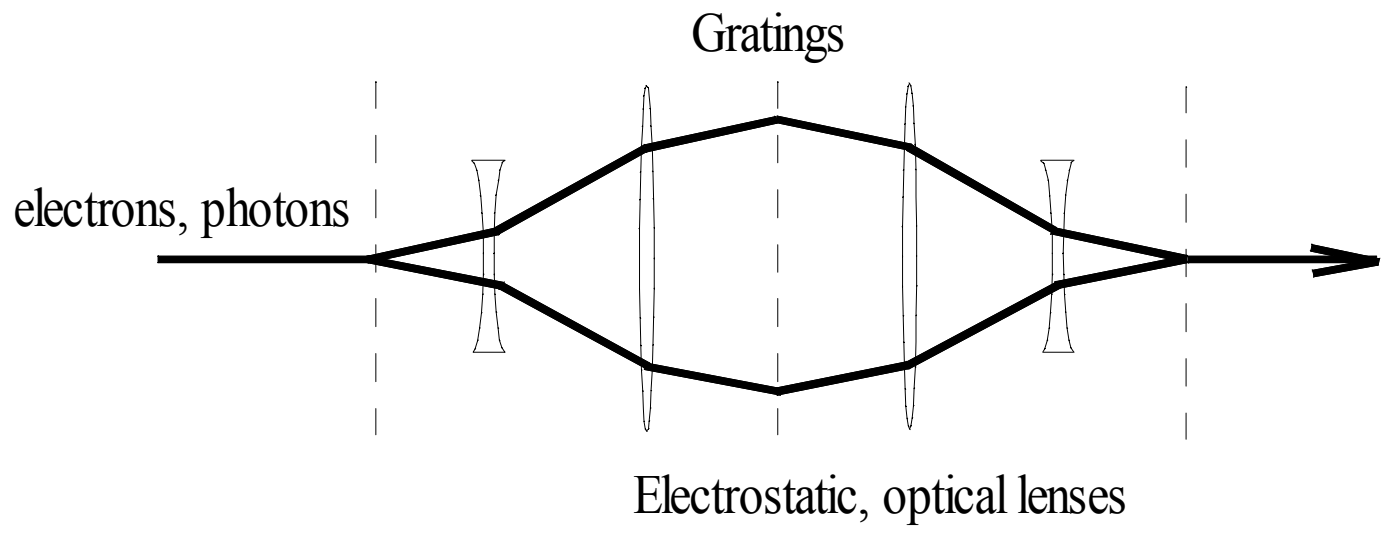

**Figure 12. The area enclosed by the two arms of a Mach-Zehnder interferometer enlarged by using lenses is shown. If this is possible is an open question.**

[Insert figure 12 about here] Subsequently a positive lens brings the diffracted beams back to their original divergence. After the second grating this process is reversed. Although the alignment of this interferometer is more difficult, a good contrast can be achieved. One can not lean to heavily on such analogies between matter optics and optics. For this case the analogy may break down because of the sharp transitions in the index of refraction as occurs for normal optics. The question if the enclosed surface of a matter interferometer can be enlarged with macroscopic fields has to our knowledge not been addressed in detail and may be limited by diffraction or highlight some other fundamental principle or may turn out to be a technical challenge.

*6.3.2 Ion interferometry*

The Schrödinger equation has traditionally been regarded as the non - relativistic equation for a spinless particle. However, in the absence of a magnetic field the Pauli equation describes two uncoupled pure spin states, which are identical to the Schrödinger equation [48]. This implies that the correct interpretation is, that particles with a fixed eigenstate of spin rather than spinless particles are described by the Schrödinger equation. This difference appears to be trivial, since in the presence of a magnetic field nobody uses Schrödinger theory but Pauli or Dirac theory, and without a magnetic field the solutions of Schrödinger and Pauli theory are identical. However, Hestenes and Gurtler [48, 49] have pointed out the deep consequences this issue has with regard to understanding the usual statement that the electron has intrinsic spin, rather than interpreting spin as a dynamical property of electron motion. Silverman sharpens the issue by pointing out that the inconsistency between the various formalisms can be tested experimentally by measuring phase shifts due to the interaction of ions with a vector potential field $\vec{A}$ [50] (figure 13). [Insert figure 13 about here] Application of minimal coupling - the replacement of the momentum $\vec{p}$ by $\vec{p}-(q/c)\vec{A}$ - leads to the Aharanov-Bohm phase shift for the spinless Schrödinger equation. Application of minimal coupling to the Dirac equation, including the spin - orbit interaction would predict different phase shifts in an Aharanov - Bohm type experiment. All tests of the Aharanov - Bohm experiment have been performed for free electrons, for which spin-orbit interaction is absent. To address this issue experimentally, an ion interferometer is necessary. For the Aharanov - Bohm effect charge is required, while spin - orbit interaction requires internal structure. Ions appear to be an ideal candidate. To build an ion interferometer a coherent beam splitter is needed. Although it seems safe to anticipate that ion interferometry will be attempted in the future it is not clear what the experimental approach will be. Either biprisms or gratings might be used. Material gratings seem less appropriate for charged particles, and with this in mind we note that no attempts to use the Kapitza-Dirac effect to diffract ions with light have been made to date.

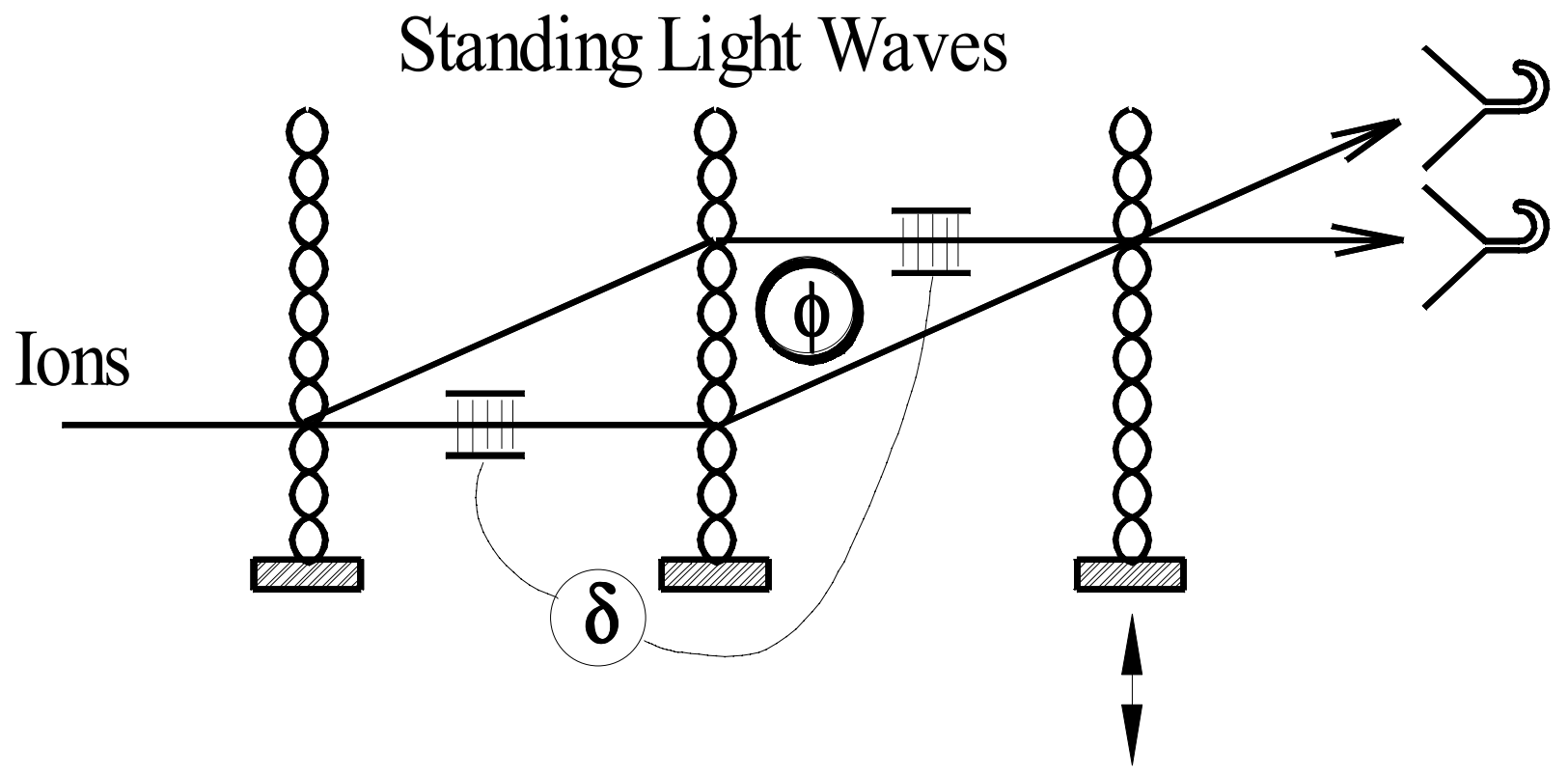

**Figure 13. (Adapted from Silverman [50]) Ion interferometry testing fundamental quantum mechanics using the Aharanov-Bohm effect. Excited ions are diffracted and pass through one or the other of two coherently oscillating fields with relative phase δ before recombining and being detected. The effect of the enclosed magnetic flux may be observed in either the ion count rate or the ion resonance fluorescence.**

### *6.3.3 Molecular interferometry*

One of the interesting things to study with a molecule interferometer would be large molecules [51]. An interesting question is, 'How large can an object be and still interfere with itself'? A first guess may be that the particle can not be larger than the spacing between adjacent lines of the used grating. This is obviously true for material gratings but not so clear for light gratings. With atoms it has been shown that particle interference persist even when the particle's physical size is larger than its de Broglie wavelength or coherence length [52]. The condition that the diffraction angle at the first grating should be large enough to include at least two adjacent rulings of the second grating, leads directly to a condition for the transit time between the gratings, $\tau = \rho s^5/h$, where $\rho$ and $s$ are the density and the size of the particle [53]. Typical transit times for atomic or molecular beam experiment are $\tau = 10^{-4}$ - $10^{-5}$ s. This means that a molecule should not contain more than $10^3$ atoms. One interesting problem that could be addressed for molecules smaller than this size is that of dephasing due to scattering of thermal photons. This dephasing or loss of coherence is an important part of our present understanding of mesoscopic physics [54]. The large excitation energy for ground state atoms as compared to molecules would not allow such a study. It also seems worth while to point out that the above condition for transit times coincides with the condition for Bragg scattering, offering the alternative of studying dephasing with a single grating, reducing the experimental complexity considerably. In such an experiment the shape of the diffraction pattern after one grating could be monitored as a function of the apparatus temperature. In this respect (and others) the recent diffraction of $C_{60}$ molecules with a material grating is a breakthrough [55]. The many internal degrees of freedom of this massive molecule allow interaction with the environment and thus provide a a means to study decoherence. Arndt *et al.* [55] even wonder if it would be possible to use optical diffraction structures to study quantum interference of small viruses.

Molecular interferometry is not only exciting in view of decoherence studies, but interferometers with heavy particles such as molecules provide one of the very few techniques where the influence of gravity on the quantum mechanical system can be measured [56]. In the first experiment of this kind the enclosed surface by the two arms of a neutron interferometer was rotated with respect to the earth gravitational field to demonstrate the equivalence principle at the quantum mechanical level. Another interesting candidate for such an experiment would be a chiral molecule, which provides a direct parity test of the principle of equivalence [39], which answers the question whether a microscopic left-handed screw and right-handed screw fall with the same acceleration. Matter interferometry is not limited to merely observe the effect of gravity on a quantum mechanical system, it can also provide tests where quantum-gravity plays a roll [57]. The basic idea is that space-time fluctuations on the Planck-scale would reduce the contrast in the interference pattern of a matter interferometer. Particles moving in both the interferometer arms function as clocks that upon interfering with each other compare their relative time. The more mass the particle has, the faster the clock runs, and the more chance one has to detect small fluctuations. Large, heavy molecules would thus provide an excellent candidate for such experiments.

**Acknowledgements**
This work was supported by the Research Corporation. The author would like to thank Markus Oberthaler and Ernst Rasel for many illuminating discussions and Dan Freimund for preparing several figures.

Herman Batelaan recently started a new laboratory in the AMO group at the University of Nebraska-Lincoln and is currently working on matter optics with intense laser light. He studied physics at the University of Leiden, receiving the degree of Drs. in 1987 and continued his studies at the University of Utrecht receiving the Ph. D in 1991. After a stay in the group of Harold Metcalf at SUNY at Stony Brook, he worked as a Lise Meitner Fellow at the University of Innsbruck in the group of Anton Zeilinger where it became clear that matter interferometry allowed the study of many facets of quantum mechanics. Visits to the University of Nebraska and the Technological University of Eindhoven followed. His main research interest is studying the foundations of quantum mechanics.